\documentclass[aps,eqsecnum,showpacs,twocolumn]{revtex4}
\usepackage{graphicx}%
 
\usepackage{amsmath}%
 
\usepackage{amsfonts}%
\usepackage{amssymb}
 
\usepackage{bm}

\begin{document}
\draft

\preprint{not yet sub}

\title{Atoms in double-$\delta$-kicked periodic potentials: chaos with long-range correlations}

\author{P.H. Jones, M. Stocklin, G. Hur, and T. S. Monteiro}

\affiliation{Department of Physics and Astronomy, University College London,
Gower Street, London WC1E 6BT, U.K.}

\date{\today}

\begin{abstract}
We report an experimental and theoretical study of the dynamics of cold atoms 
subjected to closely-spaced pairs of pulses in an optical lattice. The experiments
show the interplay between fully coherent quantum dynamics and a novel 
momentum-diffusion regime: for all previously-studied $\delta$-kicked systems, 
chaotic classical dynamics
shows diffusion with short-time (2 or 3-kick) correlations;  here,
chaotic diffusion combines with new types of long-ranged `global' correlations, between all kick-pairs,
 which control transport through trapping regions in phase-space. Analytical formulae are presented 
 and, with quantum localization, are used to analyse the experiments. 
\end{abstract}


\maketitle

The `$\delta$-kicked particle' ($\delta$-KP) is one of the most studied experimental and 
theoretical paradigms of classical Hamiltonian chaos. A particle is kicked periodically
by a sinusoidal potential $V(x,t)= -K \cos x \sum \delta(t-n)$. For sufficiently large $K$,
 the motion is fully chaotic and the energy grows diffusively. 
Of particular interest in recent years has been the theoretical \cite{Casati,Fish} and 
experimental \cite{Raizen} investigation of the suppression of the 
classical diffusive process in its quantum counterpart,  the quantum 
$\delta$-kicked particle 
($\delta$-QKP). This phenomenon is generally termed  `Dynamical Localization' (DL).
For $K > \sim 1 $  the diffusive growth of the classical energy  is no longer bounded 
by phase-space barriers (tori) so persists indefinitely : for an ensemble of
classical particles, the energy  $E=<\frac{P^2}{2}> = \frac{1}{2}D t \simeq \frac{K^2}{4}t$ for all $t$.
 The corresponding quantum energy
grows  only up to a timescale $t^* \sim D/ \hbar^2$ \cite{Shep}
and saturates at a value $ <P^2>_{t \to \infty} \sim  Dt^*$. 
The Texas experiments  showed that cesium atoms
in  periodically pulsed  waves of light were an  ideal test-bed for quantum chaos.
A broad range of interesting physical regimes were subsequently
investigated: controlled decoherence
\cite{Raizdeco}, quantum accelerator modes \cite{Darcy} and delocalization induced by
non-periodic kicking \cite{Garreau}. 

The chaotic diffusion is however not entirely uncorrelated \cite{Lich}
and there are corrections which have now been experimentally observed \cite{Anom} due  
to 2-kick and 3-kick correlations. For example, a 2-kick correction
 appears because the ensemble averaged
value for the correlation between the impulse at the $n-th$ kick and 
that experienced 2-kicks later,   $C_2= <V'(x_n)V'(x_{n+2})>$ is generally non-zero,
see \cite{Lich,Rech}. 
In \cite{Jonck,Jon1} it was further shown that,
if the pulses are unequally spaced, the 2-kick corrections yield a 
local (in momentum) correction to the diffusion. In that case, $2E(P_0,t)= D(P_0, t)\ t$ : 
for unequal kick spacings, both  the diffusion rate and hence the energy,
depend non-trivially  on time and the 
relative initial momentum, $P_0$, between the atoms and the standing wave of light.

In this work we report the first experimental and theoretical study
  of the $2\delta$-kicked particle ($2\delta$-KP):
a cloud of cesium atoms is exposed to a periodic sequence of closely spaced 
$pairs$ 
 of kicks. At the outset, one might expect that the diffusive behaviour here
could be analysed within the framework used in \cite{Anom,Jonck}, 
of diffusion with correlations between short sequences of 2 or 3 kicks, whether 
local or otherwise. 
However this approach
fails to explain the experimental results. Further investigation showed that
 chaotic diffusion in the $2\delta$-KP was rather different from that seen in 
 previously studied kicked systems. For the $2\delta$-KP one finds
 new corrections, which appear in families
  correlating all kicks democratically. These corrections
are individually very weak, but become numerous with time and accumulate to
 eventually dominate the diffusive process.
Moreover, these `global' correlations can be associated with specific physical phenomena,
namely the escape from and through `trapping' regions in phase-space. 
 We have identified  diffusive regimes associated with three types of correlation:
one is an ordinary 1-kick correlation, $C_1$, the other two are new and are 
families with `global' terms. 
 The experimental behaviour depends
 on which correlations dominate at the point when the quantum dynamics suppresses the diffusion.

\begin{figure}[ht]
\includegraphics[height=2.5in,width=2.in]{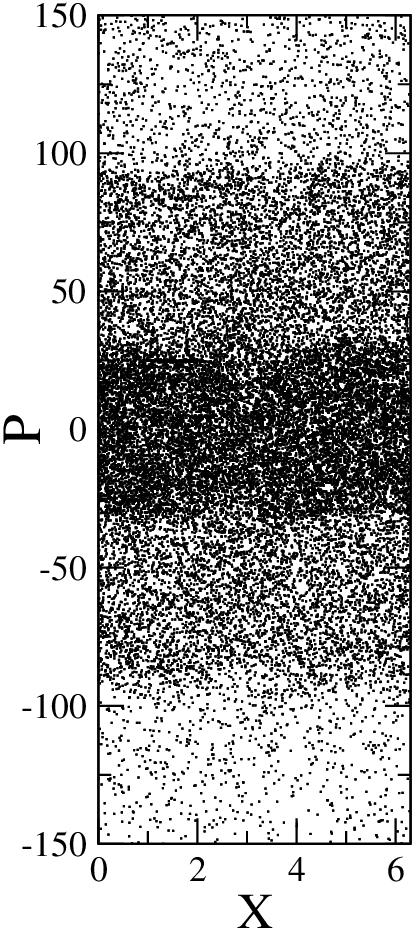}
\caption {Surface of Section diagram for an ensemble of particles,
with $K=7$ and $\epsilon=0.1$, prepared with initial momentum $P_0=0$ at $t=0$.
The SOS illustrates the trapping of trajectories in
phase space regions for which the momenta $p \epsilon \simeq \pm (2n+1) \pi$.}
\label{Fig.1}
\end{figure}
The experimental apparatus is essentially like that described in \cite{Jon1} and
consists of a cloud of cesium atoms collected in a standard 6-beam
MOT and cooled in an optical molasses to a temperature of $6\mu$K.
The sinusoidal potential $V(x,t)$ is formed by two
counter-propagating laser beams incident on the cloud with
parallel polarisations. These are derived from a Titanium Sapphire
laser, have an intensity of $4\times 10^3I_{sat}$ ($I_{sat}$
=1.12mWcm$^{-2}$, the saturation intensity) in each beam and are
detuned $2000\Gamma$ ($\Gamma = 2\pi \times 5.22$MHz, the natural
linewidth) below the D2 transition on cesium. The potential is
switched on using acousto-optic modulators (AOMs) to create pulses
as short as $t_p = 300$ns, and each beam is controlled by a
separate AOM so that a frequency difference $\Delta f$ may be
imposed upon the two beams and the potential moves with constant
velocity in the laboratory frame. In this way we may explore the
momentum dependence of the diffusion constant as in the rest frame
of the potential the atomic momentum distribution has a non-zero
mean value $P_0 \propto \Delta f$.

We now have two periods: $\tau$, which
 represents a (long) timescale between the pairs of kicks,
and $\epsilon$, which represents a much smaller time 
interval between kicks in each pair.
In the experiment, $\tau =  9.47\mu$s, while 
 five different separations of the closely-spaced pair,
in the range $0.44\mu$s to $1.48\mu$s were used: 
$\epsilon= 0.047, \ 0.063,\ 0.094, \ 0.125$ and $0.156$. For these
parameters and with the intensity and detuning as above, we have
an effective value of $\hbar=1$ while 
the kick-strength, $K = 3.3$
($\pm 10\%$ due mainly to the uncertainty in measuring the
intensity in the laser beams). Up to 100 kicks were applied before
the cloud of atoms was allowed to evolve freely in the dark for
15ms. A pair of near-resonant imaging beams were then switched on
and the fluorescence imaged on a CCD camera. From the spatial
distribution of the fluorescence the momentum distribution was
extracted and $\langle P^2 \rangle$ calculated. 

The corresponding classical behaviour of the $2\delta$-KP would be given
 by evolving a 2-kick map:
\begin{eqnarray*}
p_i = p_{i-1} + K \sin x_{i-1}  \\
x_i = x_{i-1} + p_{i} \epsilon \\
p_{i+1} = p_{i} + K \sin x_i  \\
x_{i+1} = x_i + p_{i+1} \tau \\
\end{eqnarray*}
It is instructive to begin by considering the classical evolution
 of an ensemble of particles all initially
at time $t=0$,
with momentum $p=P_0$ for which $P_0 \epsilon = (2n+1)\pi$ and $n=0,1,..$. 
These particles experience a
kick $K \sin x_{i-1}$ followed by another at $\simeq K \sin (x_{i-1}+\pi)$ which in
effect cancels the first. Conversely in the case $P_0 \epsilon = 2n \pi$ a series of
near-identical kicks produces initially rapid energy growth. Other $P_0$ produce
intermediate behaviour. This behaviour follows from the unsurprising fact that
 the presence of the short time interval,
$\epsilon$, results in a non-zero kick-to-kick correlation $C_1$. In contrast, 
for the Standard Map and $all$ other previously studied atomic $\delta$-kicked
 systems, $C_1=0$. 

\begin{figure}[ht]
\includegraphics[height=3.5in,width=3.in]{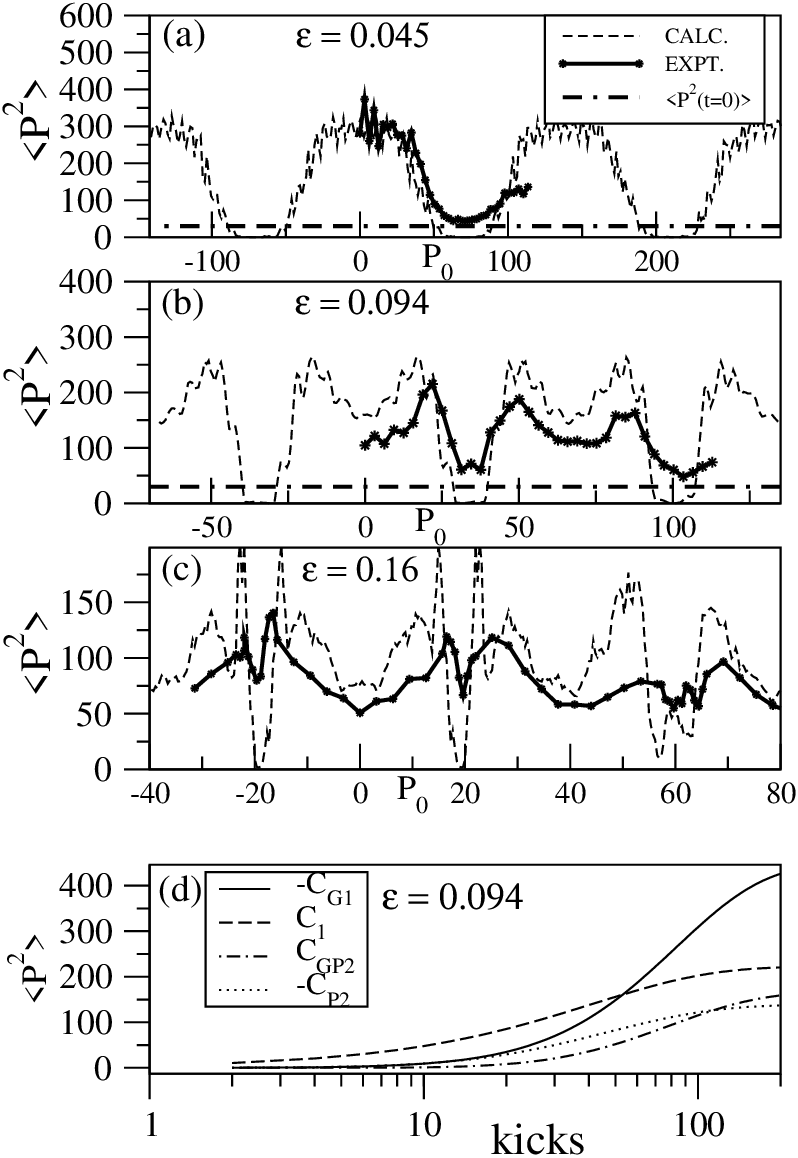}
\caption { Experimental results for $2\delta$-KP realisation with cesium atoms.
 Each data point (star) shows the energy absorbed (after 100 kicks, $K=3.3$, $\hbar=1$)  
by a cloud of atoms with average momentum $p=P_0$ (relative to
the optical lattice) at initial time, $t=0$. 
With increasing $\epsilon$, we see the minima (maxima) in the energy 
flip into maxima (minima) as the `global' correlation family $C_{G1}$
gradually  overtakes the nearest-neighbour correlation $C_1$.
The dashed lines represent a classical simulation using
 100,000 particles all with momenta $=P_0$ at $t=0$, and $K$ within
 the range $3.3 \pm 10 \%$. {\bf(a)} $t^* << t_1 \simeq 1/(K\epsilon)^2$.
 Regime dominated by the one-kick correlation $C_1$.
Atoms prepared near the trapping regions $(P_0 \epsilon \sim (2n+1)\pi)$ remain trapped.
Results follow closely the formula $<P^2> \simeq K^2 T/2 (1 + \cos P_0\epsilon)$.
{\bf(b)} $t^* \sim 1/(K\epsilon)^2$. Regime for which $C_1$ competes with the `global'
correlation family $C_{G1}$. This begins to expose the inverted peaks of
 the Poisson correlation terms $C_P$, which  determine the
trapping very close to the resonant condition $(P_0 \epsilon = (2n+1)\pi)$.
{\bf(c)} $t^* > 1/(K\epsilon)^2$. Regime dominated by
$C_{G1}$, but sharp inverted peaks due to $C_P$ are still visible.
(d) Competition between $C_1$, $C_{G1}$ and leading order Poisson terms. At short times, 
the global correlations are negligible, but at later times, the accumulation
of (individually weak) global correlations overtake the corresponding $C_1$ and $C_{Pm}$ terms: 
when the global terms become dominant, all the atoms have escaped from the trapping regions. }
\label{Fig.2}
\end{figure}

We show, in
 Fig.\ref{Fig.1}, Surface of Section plots obtained from an ensemble intially
with $P_0=0$, but randomly distributed in position $x$,
 for $K=7$, $\epsilon=0.1$. It is clear that, though 
the phase-space here is fully chaotic,
trajectories `stick' at the values of  $p \epsilon \simeq \pm (2n+1)\pi$. 
By calculating $C_1$ explicitly, after $N$ $pairs$ of kicks, one can begin a more precise analysis:
\begin{eqnarray}
 C_1(N,P_0) = K^2 \cos P_0\epsilon [J_0(K\epsilon)-J_2(K\epsilon)] \nonumber\\
              \sum_{j=1}^{N} (J_0(K\epsilon))^{2j-2}
\label{eq1}
\end{eqnarray}
Physical time is $t=N(\tau+\epsilon)$. We can re-scale variables so $\tau =1$ and $t \simeq N$. 
The ensemble-averaged energy $E$ of the atom cloud at time $t$, corrected by
Eq.\ref{eq1} would be given by
$2E =<(P_t-P_0)^2> \simeq {K^2}T/2 + C_1(t,P_0)$ (where $T=2N=2t$).  
It is easily shown that for short times $C_1$ grows linearly in time, while for longer times
it saturates to a constant value, after a time $t_1 \sim \frac{10}{(K\epsilon)^2}$.
On this time-scale, the kick-to-kick contributions decay to zero.
Since $C_1$  is a single correlation term, analogous to those studied 
in \cite{Jonck} we have here been able to simply extend the usual
analysis of \cite{Rech,Jonck}.
For small $K\epsilon$, $J_2(K\epsilon) \simeq 0$, so for short times
($t << t_1$) we can
write for the energy, $2E =<(P_t-P_0)^2> \simeq K^2 T/2 [1 + \cos P_0\epsilon] $.
The validity of this formula for $t<< t_1$ is seen in the experimental data in
Fig.\ref{Fig.2}(a). 

 Fig.\ref{Fig.2} show the energy absorbed by the 
 cesium atoms, as a function of $P_0$. Three separate values of $\epsilon$
were considered. For each $\epsilon$, the energy was measured, after $100$ kicks (but here $t^* \sim 40$)
 for many values of $P_0$.
In every case, at
 time $t=0$, the atom cloud had energy $<(P_{t=0}-P_0)^2> \simeq 30$, indicated by the horizontal
dot-dashed line.

 The experimental data of Fig.\ref{Fig.2}(a) is the 
most straightforward to 
understand: here, $t_1 \sim 450$, so $t^* << t_1$ so the energy absorption was arrested in
the regime $t<< t_1$.  An ensemble
of classical particles, initially at momenta $P_0 \epsilon \simeq \pm (2n+1)\pi$, 
absorbs no energy while for $P_0 \epsilon \simeq \pm 2n\pi$, energy absorption
is maximal.  

Fig.\ref{Fig.2}(c), on the other hand, was the most surprising.
It corresponds to a regime
$t^* > t_1$.  It shows a clear
 reversal of the behaviour seen in Fig.\ref{Fig.2}(a): atoms initially prepared
at or near the momentum-trapping regions end up with more energy than
those prepared in the enhanced momentum diffusion regions. In fact, atoms
which are prepared furthest from the momentum trapping regions,
absorb the least energy. 
\begin{figure}[htb]
\includegraphics[height=3.0in,width=3.in]{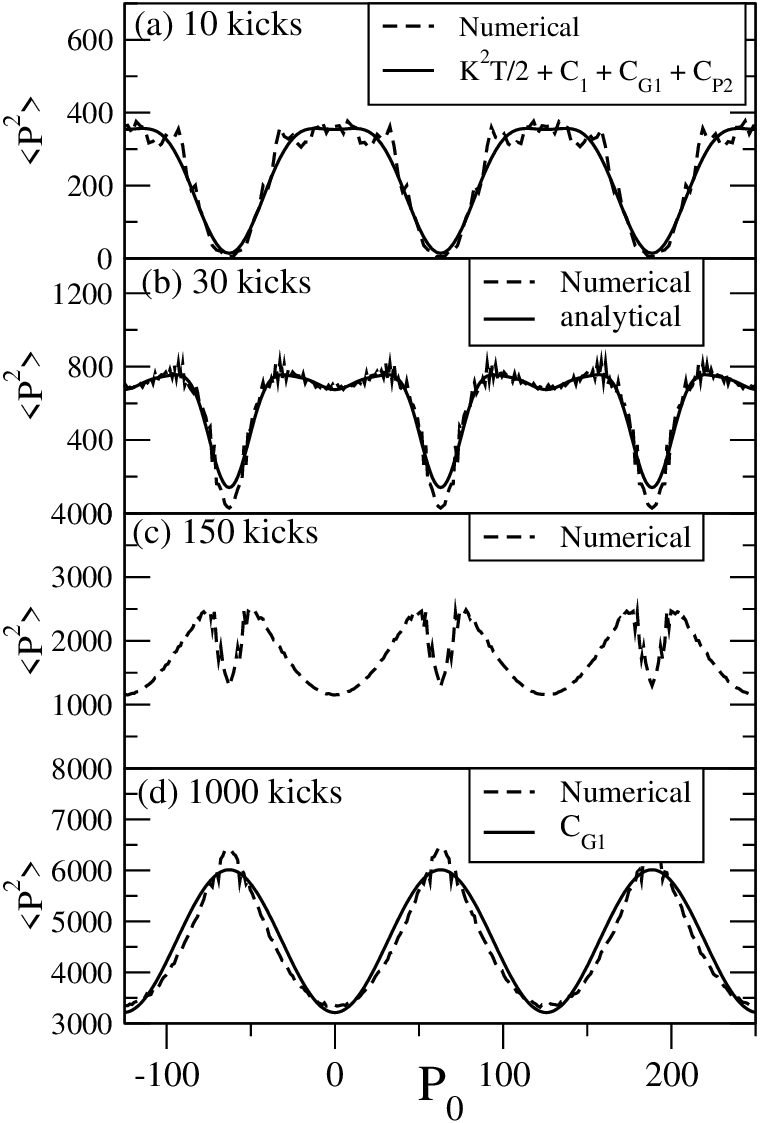}
\caption { The figure plots the energy absorbed by an ensemble of
100,000 classical particles, all with $p=P_0$ at $t=0$,
as a function of initial momentum (dashed line). These numerics are
superposed with analytical formulae obtained for the diffusion (solid
line). $K=7$ and $\epsilon=0.05$ in every figure, but the number of kicks, $2N$, varies.
Note the behaviour in the trapping region $P_0 \simeq \pi/\epsilon \simeq 60$:
the energy absorbed is a  minimum at short times but turns into a maximum
at long times.
(a) Results at short time ($10$ kicks), dominated by the one-kick correlation
 $C_1$.
(b) Results at a time for which $C_1$ competes with the `global'
de-phasing family $C_{G1}$. This exposes the `Poisson-sum-formula' corrections: 
a family
of corrections of the form $C_{Pm} \propto \cos mP_0 \epsilon$, ($m \geq 2$)
combine to give
a series of  inverted peaks in the trapping regions (c) All three corrections
$C_P, C_1$ and $C_{G1}$ compete (d) At $t > 500$ the longest lived correction,
$C_{G1}$  dominates. The Poisson corrections
have changed sign:  the inverted peaks of (c) (indicating trapping),
 have `flipped over' to give a series of positive
peaks which give the maxima a `pointed' appearance. The counter-intuitive result,
that particles initially prepared in the momentum-trapping regions will be the
ones which will acquire the most energy, is evident.
 }
\label{Fig.3}
\end{figure}

This counter-intuitive observation motivated a
 more  careful look at the mathematical model of the
diffusive process of the $2\delta$-KP. This exposed a collection of terms 
involving averaging products of the form $\sin x_i \sin x_{\mu}$, where $\mu <i$
but is otherwise arbitrary; while $i$ represents any of the $second$ kicks in
the pair. For each $\mu$ we obtain a correction of the form
$4K^2 \cos P_0\epsilon \ J_1^2(K\epsilon) \sum_j (J_0(K\epsilon))^{2j-3}$. These
terms are negligibly small ($O(K\epsilon)^2$  relative to $C_1$). However
since we sum over all $\mu < i$ their numbers accumulate with time and the net
contribution of this `global' correlation family is:
\begin{eqnarray}
 -C_{G1}(N,P_0) = 4K^2 \cos P_0\epsilon \ J_1^2(K\epsilon)\nonumber\\
\sum^{N}_{j=1} (j-1)(J_0(K\epsilon))^{2j-3}.
\label{eq2}
\end{eqnarray}

It is easily shown that, though negligible at short times, this term grows
quadratically at small $t$ and eventually overtakes $C_1$. It has opposite sign
 to $C_1$; we interpret it as a term which reflects the gradual de-phasing of the
 resonant effects of $C_1$ -such as the kick-cancellation/trapping
regions with  $P_0 \epsilon \sim (2n+1)\pi)$.
 At long times the behaviour is dominated by $C_{G1}$ and hence we see that the
 energy absorption is $maximal$ 
 for particles prepared near $P_0 \epsilon \simeq \pm (2n+1) \pi$.

Fig.2(b) corresponds to a particularly interesting regime, where $C_1 \sim C_{G1}$.
The  $\cos P_0 \epsilon$ correction is partly cancelled,
 exposing a series of  narrow dips
 in the energy. The origin of these dips is in a series
 of terms $C_{Pm} \propto \cos mP_0\epsilon$. When summed these produce 
behaviour reminiscent of the Poisson sum formula
 $\sum_m (-1)^m \cos mP_0\epsilon =  \sum_n \delta(P_0\epsilon-(2n+1)\pi)$. The amplitudes of the
$C_{Pm}$ terms vary with time and only a finite number of harmonics
($m < ~10$ typically) contributes at
any given time. Hence we get a series of broadened peaks.
 Nevertheless, for this reason, we term these corrections the Poisson term,
$C_P = \sum C_{Pm}$ and where:
\begin{eqnarray}
 C_{Pm}(N,P_0) \propto  K^2 \cos mP_0\epsilon F_m(t)\prod_{n=1}^{m-1} J_1^2(nK\epsilon).
\label{eq3}
\end{eqnarray}
 $F_m(t)$ is a time function which grows as $\sim t^m$. Though these terms are small ($O(K\epsilon)^{2m-2}$)
they will contribute when $t^m (K\epsilon)^{2m-2} \sim 1$. Each of these terms 
also has a partner `global' family of opposite sign, $C_{GPm}$ ($O(K\epsilon)^{2m}$), which increase as $\sim t^{m+1}$.
 Fig.2 (d) shows the behaviour of the $m=2$
Poisson terms in comparison to $C_1$ and $C_{G1}$. As $m$ increases terms become less significant and for
each $m$ the global correlation always dominates $C_{Pm}$ at long times. There are  additional, even higher-order, 
$\cos nP_0 \epsilon$ terms ($n\geq 1$) of similar form to
$C_{Pm}$ and $C_{GPm}$ above and a group of terms involving products of the form $\sin^2 x_i$ which may
contribute, particularly to $C_{P2}$.

In Fig.\ref{Fig.3} we compare numerical (classical) calculations at $K=7, \ \epsilon=0.05$,
 with the behaviour predicted
by the correlations. In Fig.\ref{Fig.3}(a) we look at short times and see that we can
accurately match the energy absorption by including only the three lowest order diffusive corrections,
ie those which increase linearly or quadratically  in time ($C_1, \ C_{G1}, \ C_{P2}$).
At later times ($t < 50$ or so), we obtain good agreement by including all the above terms up 
to order $J_1^{10}$ and $m=4$: the inverted peaks corresponding to the trapping regions are quite well reproduced,
as seen in Fig.\ref{Fig.3}(b).
At extremely long times, we can obtain good results simply from  the leading
global family $C_{G1}$. We also clearly see, in Fig.\ref{Fig.3}(d), the inversion of the behaviour seen at earlier times:
atoms prepared within the trapping regions give positive peaks, since $C_{GPm}$ dominates.
 The $0-th$ order term
is $<P^2> \simeq K^2T/2$ at very short times, then gradually slows down to an
asymptotic value $\simeq K^2T/16$ (obtained numerically) at long times. The analytical curves
in Fig.3 are shifted vertically by a constant amount.

In conclusion, we have presented and analysed an experimental realisation of a 
$2\delta-KP$ and shown that it corresponds to a type of diffusion quite different from the Standard Map.
This is a rich and complex system and many questions remain open. On the classical side, certain 
aspects of the interactions and lifetimes of the different types of correlations are not yet well understood. 
Although we have a generic handle on this system with the
diffusion correlations, further insight might be gained by detailed knowledge of classical trajectories.
For  lower values
of  $K\epsilon$, a thin band of islands and eventually unbroken tori appears first around
$p \epsilon \simeq \pm (2n+1)\pi$. At this point, clearly, there will be no
escape through the trapping regions and beyond the very shortest times 
($t_1 << 1/(K\epsilon)^2$)  the diffusive correlation approach
would fail everywhere. We expect regions permeated by  broken phase-space barriers 
to persist at the parameters such as eg. Fig.\ref{Fig.1} and to account for the trapping regions.

The quantum behaviour is also not fully understood. Though clearly some key features can 
be modelled qualitatively in Fig.2 by adjusting the number of kicks and $K$ (by up to 
$10\%$ in the classical numerics of Fig.2), the quantum dynamics will involve tunnelling and
dynamical localization (DL) effects which can introduce substantial differences with the classical behaviour.
Further investigation of the DL in this system is necessary: its Floquet states are localised,
like in the standard $\delta-KP$. But in the latter, localization lengths, $L$, are quite uniform
with exponential localization and $L \sim \frac{D}{\hbar}$, while in the $2\delta-KP$, values of $L$ 
can vary by about 3 orders of magnitudes for the parameters considered here \cite{gohur}.

Although the $t << t_1$ experimental regime (seen in Figs. 2(a) and 3(a))
is in some sense the least surprising in terms of the diffusive process,
 it is worth noting its potential for atomic manipulation. 
Interest in the $2\delta$-KP experiment was initially 
stimulated by its potential applications
in manipulating atoms in devices like an atom `chip' \cite{Hinds}. In \cite{Jonck} it was
proposed that a local diffusion rate $D(P_0)$, could be exploited for `filtering' cold atoms 
according to their velocity. For selected $P_0$ the atoms could pass the device unperturbed,
while other momenta would absorb a substantial amount of energy and would
be dispersed. The $2\delta$-QKP  showed a far stronger experimental signature
than the system in \cite{Jon1,Jonck} which relied on a two-kick, $C_2$, correlation
(note that $C_2 \neq C_{P2}$). A much stronger
 velocity-selective effect, due to the $C_1$, correlation is  seen for the 
$2\delta$-QKP. The inverted
peaks of the $C_p$ correlations could also be used to select a narrow band of velocities
with $P_0 \simeq \pi/\epsilon$.

The authors thank Thibaut Jonckheere for helpful discussions. 
This work was supported by the EPSRC.

\end{document}